\documentclass[
    ,final            
  ]
  {aipproc}

\layoutstyle{8x11single}

\begin{document}

\title{Exploring the host environments of long-duration gamma-ray bursts}


\author{Emily M. Levesque\footnote{Einstein Fellow}}{
  address={CASA, Dept. of Astrophysical and Planetary Sciences, University of Colorado 389-UCB, Boulder, CO 80309}
}




\begin{abstract}
We have conducted the first dedicated spectroscopic survey of long-duration gamma-ray burst (LGRB) host galaxies at $z < 1$, and use these observations along with data from the literature to determine a wide range of ISM properties and a statistically robust mass-metallicity relation. LGRBs have been proposed as possible tracers of star formation at high redshift; however, such an association is dependent on a thorough understanding of the relationship between LGRB progenitors and their host environments. In particular, the metallicity of LGRB host galaxies has become a matter of hot debate in recent years. We conclude that LGRBs do exhibit a general trend toward lower-metallicity host galaxies, but also detect several high-metallicity hosts in our sample. We have also compared the energetic and environmental properties of the LGRBs in our sample, and find no statistically significant correlation between host metallicity and isotropic or beaming-corrected gamma-ray energy release. This is at odds with previous theoretical and observational predictions of an inverse correlation between gamma-ray energy release and host metallicity, and demonstrates that the complex role of metallicity in LGRB progenitor formation still remains unclear.
\end{abstract}

\maketitle


\section{Introduction}
Long-duration ($>$2 s) gamma-ray bursts (LGRBs) are among the most energetic events observed in our universe, and are typically associated with the core-collapse deaths of unusual massive stars \cite{W93}. As a result, they are often cited as potential tracers of the star formation and metallicity history of the early universe \cite{Bloom02,Fynbo07,Savaglio09}.

However, in recent years several studies of these events' host galaxies have suggested that LGRBs may occur preferentially in low-metallicity environments, which would threaten their utility as unbiased star formation tracers. Much of this work on LGRB host metallicities has focused on comparison with the general star-forming galaxy population in the context of the standard luminosity-metallicity {\it L-Z} relation for star-forming galaxies \cite{Stanek06,Kewley07,Modjaz08,Levesque2010a}. Such comparisons rely on luminosity as a proxy for stellar mass. Since luminosity is also dependent on the star formation rate and metallicity of a host galaxy, performing these same comparisons using the mass-metallicity ({\it M-Z}) relation offers a more direct means of isolating the effects of metallicity in the galaxy samples and probing its role in LGRB host galaxies.

In addition, the physical mechanism driving a potential metallicity bias in LGRB host galaxies remains unclear. If this environmental correlation arises from a direct impact on progenitor properties (such as metallicity-dependent angular momentum), then we would expect metallicity to also show some correlation with the high-energy prompt emission properties of LGRBs. Specifically, a lower-metallicity progenitor will have a higher helium core mass and a faster rotation rate, producing a more energetic LGRB \cite{MW99}; in other words, a low-metallicity host environment should produce a LGRB with a higher energy release in the gamma-ray regime ($E_\gamma$). Several studies have investigated this potential relation, but have been limited by a lack of host galaxy data for LGRBs \cite{Stanek06,RR02,WP07}.

Here we present recent results examining the {\it M-Z} relation of LGRB host galaxies, and a comparison of their host metallicity and energetic properties. We determined stellar masses and host metallicities for a sample of 16 $z < 1$ LGRBs and compare them to other star-forming galaxy samples at similar redshifts. We also use prompt emission data from the literature to conduct a statistical comparison between LGRB host metallicities and their isotropic ($E_{\gamma, iso}$) and beaming-corrected ($E_{\gamma}$) energy release. Based on these comparisons, we consider the implications for LGRB progenitor models and our understanding of the role that metallicity plays in LGRB production.

\section{The Mass-Metallicity Relation for LGRBs}
We are conducting an ongoing uniform rest-frame optical spectroscopic survey of $z < 1$ LGRB host galaxies, using the Keck telescopes at Mauna Kea Observatory and the Magellan telescopes at Las Campanas Observatory \cite{Levesque2010a,Levesque2010b}. We have restricted our sample to confirmed long-duration bursts, selected from the GHostS archive \cite{Savaglio06} and the Gamma-Ray Burst Coordinates Network. To date, this survey has acquired spectroscopic data for 12 $z < 1$ LGRB hosts. Combined with four host galaxies that have high-quality spectroscopic data in the literature, we present a current sample of 16 LGRB host galaxies.

Our spectra were first corrected for the total line-of-sight E($B-V$), determined based on the fluxes of the Balmer lines and the standard reddening law \cite{CCM}. We then determined metallicities based on a calibration of the ([OIII] $\lambda$5007 + [OIII] $\lambda$4959 + [OII] $\lambda$3727]/H$\beta$ ($R_{23}$) diagnostic \cite{KD02,KK04}. In two cases (the host galaxies of GRB 020405 and GRB 980703), we could not distinguish whether the host metallicity lay on the ``lower" or ``upper" branch of the double-valued R$_{23}$ diagnostic calibration; we therefore performed all subsequent analyses with both metallicities for completeness. Finally, several host galaxies (the hosts of GRBs 980425, 990712, 030528, and 050824) fell on the ``turn-over" of the diagnostic, corresponding to the metallicity of log(O/H) + 12 $\sim$ 8.4. For our full sample we found an average R$_{23}$ metallicity of log(O/H) + 12 = 8.4 $\pm$ 0.3. 

We also estimated stellar masses for our LGRB host galaxies. Using the {\it Le Phare}\footnote{\url{http://www.cfht.hawaii.edu/~arnouts/LEPHARE/cfht_lephare/lephare.html}} code, we fit multiband photometry for the host galaxies \cite{Savaglio09} to stellar population synthesis models adopting a Chabrier IMF, the Bruzual \& Charlot synthetic stellar templates, and the Calzetti extinction law \cite{BC03,Chabrier03,Calzetti00}.  The fitting yielded a stellar mass probability distribution for each galaxy, with the median of the distribution serving as our estimate of the final stellar mass; for a more detailed discussion of the mass determination see Ilbert et al.\ (2009) and Levesque et al.\ (2010b). For three of the hosts in our sample (the hosts of GRBs 050824, 070612A, and 020405) we had insufficient photometric data to determine a stellar mass estimate. For our remaining sample of 13 LGRB host galaxies, we find a mean stellar mass of log($M_{\star}/M_{\odot}) = 9.25^{+0.19}_{-0.23}$.

The metallicities and stellar masses that we derived were used to construct a {\it M-Z} relation for LGRB host galaxies, which we plot in Figure 1. We find a strong and statistically significant correlation between stellar mass and metallicity for LGRB hosts out to $z < 1$ (Pearson's $r = 0.80$, $p = 0.001$). We also compare our results to two samples of star-forming galaxies in Figure 1. The nearby ($z < 0.3$) LGRB hosts are compared to data from $\sim$53,000 star-forming SDSS galaxies, while the intermediate-redshift ($0.3 < z < 1$) LGRB hosts are plotted with data from a sample of 1,330 emission line galaxies from the Deep Extragalactic Evolutionary Probe 2 (DEEP2) survey \cite{T04,JZ10}. From these comparisons, we find that most of the LGRB hosts in our sample fall below the standard {\it M-Z} relation for star-forming galaxies at similar redshifts; the differences range from $-0.10$ to $-0.75$ dex across a fixed stellar mass, with an average offset from the general star-forming galaxy population of $-0.42 \pm 0.18$ dex in metallicity.

Interestingly, while we see clear evidence of a low-metallicity trend in LGRB host galaxies, we see no evidence that there is a clear metallicity cut-off above which LGRBs cannot be formed, a conclusion at odds with previous studies of LGRB host galaxies and our current understanding of their progenitor evolution \cite{Modjaz08,Levesque2010a,Levesque2010b,Kocevski09}. This is made particularly clear in the unusual cases of the high-metallicity GRB 020819B and GRB 050826 hosts, both of which agree with the general star-forming galaxy samples to within the systematic errors. We determine a global host metallicity for GRB 050826, but in the case of GRB 020819B we are able to determine a metallicity for the precise explosion site of the GRB and find an R$_{23}$ metallicity of log(O/H) + 12 = 9.0 $\pm$ 0.1 \cite{Levesque2010c}. In addition to these two high-metallicity LGRB hosts, a high-metallicity explosion environment was recently found for the relativistic supernova SN 2009bb, with explosive properties found to be extremely similar to the supernovae that accompany nearby LGRBs \cite{Soderberg2010,Levesque2010d}. This is surprising in light of previous predictions of a low cut-off metallicity for such phenomena.
\begin{figure}
  \includegraphics[height=.4\textheight]{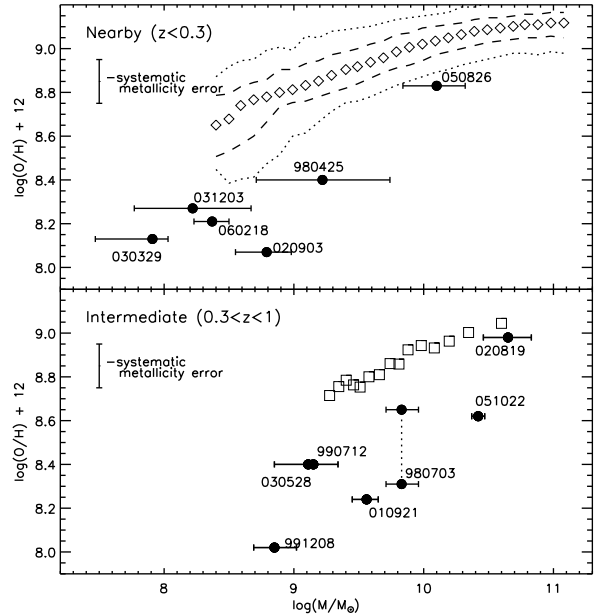}
  \caption{{\it M-Z} relation for nearby ($z < 0.3$, top) and intermediate-redshift ($0.3 < z < 1$, bottom) LGRB host galaxies (filled circles). The nearby LGRB hosts are compared to binned {\it M-Z} data for a sample of $\sim$53,000 SDSS star-forming galaxies, where the open diamonds represent the median of each bin and the dashed/dotted lines show the contours that include 68\%/95\% of the data \cite{T04}. For the intermediate-redshift hosts, we plot binned mass-metallicity data for a sample of 1330 emission line galaxies from the DEEP2 survey (open squares) \cite{JZ10}. For the $z = 0.966$ host galaxy of GRB 980703, where we cannot distinguish between the lower and upper metallicities given by the R$_{23}$ diagnostic, we plot both metallicities and connect the resulting data points with a dotted line to indicate their common origin from the same host spectrum.}
\end{figure}

\section{Energetics and Host Metallicity of LGRBs}
To compare the host metallicities and energetic properties of the LGRBs in our sample, we adopted values for $E_{\gamma, iso}$ from several different sources in the literature \cite{Amati06,Amati08,Butler07,Starling2010}. We converted $E_{\gamma,iso}$ into $E_{\gamma}$ using estimates of the jet opening angle $\theta_j$ and the convention $E_{\gamma} = E_{\gamma,iso} \times 1-cos(\theta_j)$ \cite{Frail01,Levesque2010e}. For this work we exclude GRB 070612A because of insufficient energetics data, and include the recent nearby GRB 100316D \cite{Starling2010,Chornock2010}.

We plot our results in Figure 2, comparing host metallicity to redshift, $E_{\gamma,iso}$, and $E_{\gamma}$. We include the redshift comparison to highlight any potential correlation that may appear as an artifact of metallicity evolution with redshift. We find that there is no statistically significant correlation between metallicity and redshift; more interestingly, we draw the same conclusion when comparing metallicity and $E_{\gamma,iso}$ {\it or} $E_{\gamma}$. This result is at odds with the inverse correlation predicted by previous studies (\cite{Stanek06, MW99}), and appears to demonstrate that even when taking beaming effects into account, there is no evidence for a relation between host metallicity and gamma-ray energy release in LGRBs.

\begin{figure}
  \includegraphics[height=.3\textheight]{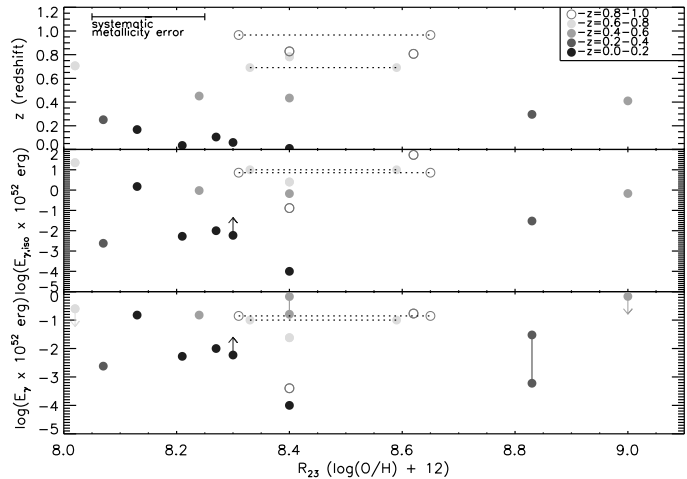}
  \caption{Metallicity vs. redshift (top), $E_{\gamma,iso}$ (center), and $E_{\gamma}$. The hosts are separated into redshift bins in order to better illustrate redshift effects.  Two hosts with both lower and upper-branch R$_{23}$ metallicities (the hosts of GRB 020405 at $z = 0.691$, and GRB 980703 at $z = 0.966$) are shown as lower and upper data points connected by dotted lines. Upper and lower limits are indicated by arrows. Hosts with both upper {\it and} lower limits on their $E_{\gamma}$ values are shown as data points connected by solid lines.}
\end{figure}

\vspace{10pt}

Based on the work described above, it appears that the physical mechanism driving this apparent metallicity trend in LGRB host galaxies currently remains a mystery. In the future, extending the comparisons shown above to include more LGRBs across a greater range of redshifts would help to improve our understanding of these enigmatic events.

\begin{theacknowledgments}
Megan Bagley, Edo Berger, Andy Fruchter, John Graham, Lisa Kewley, Alicia Soderberg, and Jabran Harus Zahid were invaluable co-authors and collaborators during this work. We are grateful for the hospitality and assistance of the W. M. Keck Observatories in Hawaii and the Las Campanas Observatory in Chile. We recognize and acknowledge the very significant cultural role and reverse that the summit of Mauna Kea has always had within the indigenous Hawaiian community, and are most fortunate to have the opportunity to conduct observations from this sacred mountain.
\end{theacknowledgments}




\bibliographystyle{aipproc}   


\end{document}